\documentclass[reprint, amsmath, amssymb, aps]{revtex4-1}
\usepackage{graphicx}
\usepackage{dcolumn}
\usepackage{bm}
\usepackage{mathrsfs}
\usepackage{hyperref}
\usepackage{epstopdf}
\usepackage{amsmath}
\usepackage{amssymb}
\usepackage{mathrsfs}
\usepackage{diagbox}

\begin{document}
\title{Lagrangian formulation for emergent elastic waves in magnetic emergent crystals}
\author{Yangfan Hu}
 \email[Corresponding author.]{huyf3@mail.sysu.edu.cn}
\affiliation{Sino-French Institute of Nuclear Engineering and Technology, Sun Yat-sen University, 519082, Zhuhai, China}

\begin{abstract}
    Magnetic emergent crystals are periodic alignment of ``particle-like'' spin textures that emerge in magnets. Instead of focusing on an individual spin or a macroscopic magnetization field, we analyze the dynamical behaviors of these novel states by taking a solid-state point of view. Based on variational principles, the basic equations for lattice dynamics of any emergent crystalline states appearing in magnetic materials is established. For small amplitude emergent elastic waves propagating in emergent crystals, the basic equations reduce to an eigenvalue problem, from which the dispersion relation and vibrational patterns for all emergent phonons are determined at the long wavelength limit.
\end{abstract}

\maketitle

\textit{Introduction}

Since the first successful experimental observation of Bloch-type 2D magnetic skyrmion crystals (SkX) in bulk MnSi\cite{1}, SkX with different chirality\cite{2}, commensurability\cite{3,4} and dimensionality\cite{5}, or even crystalline states of other spin textures such as antiskyrmion crystal \cite{6} and bi-skyrmion crystal\cite{7,8} are discovered in various kinds of magnetic materials. As emergent crystals (ECs), the collective spin excitations of these macroscopic states and related properties have been extensively studied\cite{9,10,11,12,13,14,15,16,17,18,19,20,21} due to their scientific interest and application potential in magnonics. In terms of theoretical modeling, most of the studies start by numerically solving the well-known Landau-Lifshitz-Gilbert\cite{22,23} (LLG) equation of the magnetization field, where an explicit understanding of the dynamical properties is difficult to be obtained. If we are only interested in deriving the collective excitations of these ECs around some equilibrium state, it is more convenient to take into account the periodicity of these ECs, and expand the magnetization $\mathbf{M}$ as a Fourier series\cite{24,25}, where the dynamical behaviors around the emergent crystalline state considered is described by a vibrating emergent displacement field and vibrating Fourier magnitudes. In previous studies, dynamics of the emergent displacement field\cite{17,26} and the Fourier magnitudes\cite{10} have been studied individually, while their coupling has not been considered. Based on variational principles, we establish the basic equations describing the coupling small amplitude vibration of the emergent displacement field and the Fourier magnitudes at long wavelength limit for any magnetic Ecs. Considering ECs as periodic alignment of “emergent particles”, i.e., localized field patterns instead of point masses, the equations obtained describe the lattice dynamics of these ECs, where vibration of the emergent displacement field describes lattice vibration and vibration of the Fourier magnitudes describes in-lattice vibration (i.e., vibration of the field pattern inside the lattice while keeping the lattice unchanged). Generally, the derived equations are hard to solve. Yet, for emergent elastic waves (long wavelength emergent phonons) propagation, we derive from the basic equations an eigenvalue problem which determines the dispersion relation of emergent phonons near long wavelength limit.
\\
\\
\textit{Variational principles for dynamics of the magnetization field}

The Euler-Lagrange equation for dynamics of magnetization can be derived from the principle of least action $\delta S=0$, or equivalently 
\begin{equation}
\begin{aligned}
\delta \int{L}dt=0,
\label{1}
\end{aligned}
\end{equation}
where $S$ is the action of the system, and $L={{E}_{K}}-\Phi $ is the Lagrangian of the system, with ${{E}_{K}}$ and $\Phi $ denoting respectively the total kinetic energy induced by dynamical behavior of the magnetization and the total free energy. It has been mentioned long ago that the form of ${{E}_{K}}$ is not unique. The most commonly used form derives from the Berry Phase action of a spin, which gives ${{E}_{K}}=\frac{{M}}{\gamma }\int{\dot{\varphi }\cos \theta dV}$\cite{27,28}, where ${{M}}$ denotes the average modulus of magnetization, $\gamma $ denotes the gyromagnetic ratio, and $\varphi $ and $\theta $ are Euler angles of the magnetization field. Alternatively, we can use the following variational form of ${{E}_{K}}$:
\begin{equation}
\begin{aligned}
\delta {{E}_{K}}=\frac{{{M}}}{\gamma }\int{\left( \mathbf{n}\times \mathbf{\dot{n}} \right)\delta \mathbf{n}dV},
\label{2}
\end{aligned}
\end{equation}
where $\mathbf{n}$ denotes the unit magnetization vector. Substitution of eq. (2) into eq. (1) yields the well known equation 
\begin{equation}
\begin{aligned}
\mathbf{\dot{n}}=\gamma {{\mathbf{H}}_{eff}}\times \mathbf{n},
\label{3}
\end{aligned}
\end{equation}
where ${{\mathbf{H}}_{eff}}=-\frac{1}{{{M}}}\frac{\delta \Phi }{\delta \mathbf{n}}$ denotes the effective magnetic field. If dissipation is considered\cite{23}, eq. (1) becomes 
\begin{equation}
\begin{aligned}
\delta \int{L}dt-\int{\frac{\partial W}{\partial \mathbf{\dot{n}}}}\delta \mathbf{n}dt=0,
\label{4}
\end{aligned}
\end{equation}
where $W=\frac{\alpha {{M}}}{2\gamma }{{\int{{\mathbf{\dot{n}}}}}^{2}}dV$. Substitution of eq, (2) into eq. (4) yields the Landau-Lifshitz-Gilbert (LLG) equation\cite{22,23}
\begin{equation}
\begin{aligned}
\mathbf{\dot{n}}=\gamma {{\mathbf{H}}_{eff}}\times \mathbf{n}-\alpha \mathbf{\dot{n}}\times \mathbf{n}.
\label{5}
\end{aligned}
\end{equation}
When the free energy $\Phi $ can be expressed in terms of the unit magnetization vector $\mathbf{n}$, the upper equations are directly applicable. On the other hand, if we consider the modulation of modulus of magnetization due to a change of the temperature or due to the presence of ECs, we have $\Phi =\Phi (\mathbf{M})$, where $\mathbf{M}$ denotes the magnetization vector. In this case, eq. (2) should be replaced by
\begin{equation}
\begin{aligned}
\delta {{E}_{K}}=\frac{1}{\gamma {{{{M}}}^{2}}}\int{\left( \mathbf{M}\times \mathbf{\dot{M}} \right)\delta \mathbf{M}dV},
\label{6}
\end{aligned}
\end{equation}
where ${M}=\frac{1}{V}\int{\left| \mathbf{M} \right|dV}$.
\\
\\
\textit{Fourier representation of deformable emergent crystals in magnetic materials}

Deformable ECs in helimagnets permit the following Fourier expansion of the magnetization within the Eulerian coordinates\cite{25}
\begin{equation}
\begin{aligned}
\mathbf{M}=\sum\limits_{\mathbf{l}}{{{\mathbf{M}}_{{{\mathbf{q}}_{\mathbf{l}}}}}{{e}^{\text{i}{{\mathbf{q}}_{\mathbf{l}}}[\mathbf{I}-{{F}^{e}}(\mathbf{r})]\cdot \mathbf{r}}}},
\label{7}
\end{aligned}
\end{equation}
where ${{\mathbf{q}}_{\mathbf{l}}}$ denotes the reciprocal lattice vectors of the EC. For a $d$-dimensional EC ($d=1,\ 2,\ 3$), ${{\mathbf{q}}_{\mathbf{l}}}={{l}_{1}}{{\mathbf{q}}_{1}}+{{l}_{2}}{{\mathbf{q}}_{2}}+\cdots +{{l}_{d}}{{\mathbf{q}}_{d}},$ where $\mathbf{l}={{\left[ {{l}_{1}},\ {{l}_{2}},\ \cdots ,\ {{l}_{d}} \right]}^{T}}$ is a vector of integers, and ${{\mathbf{q}}_{1}}$, ${{\mathbf{q}}_{2}}$, …, ${{\mathbf{q}}_{d}}$ are the basic reciprocal vectors. In eq. (7), $F_{ij}^{e}(\mathbf{r})=\varepsilon _{ij}^{e}+\omega _{ij}^{e}$, where $\varepsilon _{ij}^{e}=\frac{1}{2}\left( u_{i,j}^{e}+u_{j,i}^{e} \right)$ are components of the emergent strain tensor, $\omega _{ij}^{e}=\frac{1}{2}\left( u_{i,j}^{e}-u_{j,i}^{e} \right)$ are components of the emergent rotation tensor, and $u_{i}^{e}$ are components of the emergent displacement vector. Disregarding rigid translation of the EC considered, two types of deformation may occur\cite{25}: the lattice deformation described by $\varepsilon _{ij}^{e}$ and $\omega _{ij}^{e}$ which transform the undeformed wave vectors ${{\mathbf{q}}_{\mathbf{l}}}$ to $\mathbf{q}_{\mathbf{l}}^{e}\left( \varepsilon _{ij}^{e},\ \omega _{ij}^{e} \right)={{\left[ \mathbf{I}-{{\mathbf{F}}^{e}}(\mathbf{r}) \right]}^{T}}{{\mathbf{q}}_{\mathbf{l}}}$, and the in-lattice deformation described by variation of the Fourier magnitudes ${{\mathbf{M}}_{{{\mathbf{q}}_{\mathbf{l}}}}}$. 
Consider the dynamics of magnetization when the equilibrium state of the system is stabilized in an emergent crystalline state, eq. (7) becomes
\begin{equation}
\begin{aligned}
\mathbf{M}=&\sum\limits_{\mathbf{l}}\left[ {{\left( {{\mathbf{M}}_{{{\mathbf{q}}_{\mathbf{l}}}}} \right)}_{st}}+{{\left( {{\mathbf{M}}_{{{\mathbf{q}}_{\mathbf{l}}}}}(\mathbf{r},t) \right)}_{v}} \right] \\ &
\times {{e}^{\text{i}\mathbf{q}_{\mathbf{l}}^{e}\left( {{\left( \varepsilon _{ij}^{e} \right)}_{st}},\ {{\left( \omega _{ij}^{e} \right)}_{st}} \right)\cdot \left[ \mathbf{r}-{{\left( {{\mathbf{u}}^{e}}(\mathbf{r},t) \right)}_{v}} \right]}},
\label{8}
\end{aligned}
\end{equation}
where ${{\left( P \right)}_{st}}$ denotes the static value of the quantity $P$, and ${{\left( P \right)}_{v}}$ denotes time-dependent departure from ${{\left( P \right)}_{st}}$ due to dynamical behavior of $P$. As introduced before\cite{10,25}, one can take all components of ${{\left( {{\mathbf{M}}_{{{\mathbf{q}}_{\mathbf{l}}}}}(\mathbf{r},t) \right)}_{v}}$ for all possible choices of ${{\mathbf{q}}_{\mathbf{l}}}$, and construct a large vector describing the vibration of all Fourier magnitudes ${{\left( {{\mathbf{M}}^{q}}(\mathbf{r},t) \right)}_{v}}$. Eq. (8) shows that the dynamic behavior of any magnetic EC around an equilibrium state is determined by the vibrating emergent displacement field ${{\left( {{\mathbf{u}}^{e}}(\mathbf{r},t) \right)}_{v}}$ and the vibrating Fourier magnitudes ${{\left( {{\mathbf{M}}^{q}}(\mathbf{r},t) \right)}_{v}}$. 
\\
\\
\textit{Basic equations of emergent elastic waves in magnetic emergent crystals}

The basic equations of emergent elastic wave propagation in magnetic ECs can be derived by substituting eq. (8) into eq. (1), which gives
\begin{equation}
\begin{aligned}
-\frac{d}{dt}\left[ \frac{\partial L}{\partial {{\left( {{{\mathbf{\dot{u}}}}^{e}} \right)}_{v}}} \right]+\frac{\partial L}{\partial {{\left( {{\mathbf{u}}^{e}} \right)}_{v}}}-\sum\limits_{i}{\frac{d}{d{{r}_{i}}}\left[ \frac{\partial L}{\partial {{\left( \mathbf{u}_{,i}^{e} \right)}_{v}}} \right]}=0,
\label{9}
\end{aligned}
\end{equation}
\begin{equation}
\begin{aligned}
-\frac{d}{dt}\left[ \frac{\partial L}{\partial {{\left( {{{\mathbf{\dot{M}}}}^{q}} \right)}_{v}}} \right]+\frac{\partial L}{\partial {{\left( {{\mathbf{M}}^{q}} \right)}_{v}}}-\sum\limits_{i}{\frac{d}{d{{r}_{i}}}\left[ \frac{\partial L}{\partial {{\left( \mathbf{M}_{,i}^{q} \right)}_{v}}} \right]}=0.
\label{10}
\end{aligned}
\end{equation}
More conveniently, we have from eq. (2)
\begin{equation}
\begin{aligned}
\frac{\delta {{E}_{K}}}{\delta {{\left( {{\mathbf{u}}^{e}} \right)}_{v}}}-\frac{\partial \Phi }{\partial {{\left( {{\mathbf{u}}^{e}} \right)}_{v}}}+\sum\limits_{i}{\frac{d}{d{{r}_{i}}}\left[ \frac{\partial \Phi }{\partial {{\left( \mathbf{u}_{,i}^{e} \right)}_{v}}} \right]}=0,
\label{11}
\end{aligned}
\end{equation}
\begin{equation}
\begin{aligned}
\frac{\delta {{E}_{K}}}{\delta {{\left( {{\mathbf{M}}^{q}} \right)}_{v}}}-\frac{\partial \Phi }{\partial {{\left( {{\mathbf{M}}^{q}} \right)}_{v}}}+\sum\limits_{i}{\frac{d}{d{{r}_{i}}}\left[ \frac{\partial \Phi }{\partial {{\left( \mathbf{M}_{,i}^{q} \right)}_{v}}} \right]}=0.
\label{12}
\end{aligned}
\end{equation}
When damping is considered, eqs. (11-12) transform to
\begin{equation}
\begin{aligned}
\frac{\delta {{E}_{K}}}{\delta {{\left( {{\mathbf{u}}^{e}} \right)}_{v}}}-\frac{\partial \Phi }{\partial {{\left( {{\mathbf{u}}^{e}} \right)}_{v}}}+\sum\limits_{i}{\frac{d}{d{{r}_{i}}}\left[ \frac{\partial \Phi }{\partial {{\left( \mathbf{u}_{,i}^{e} \right)}_{v}}} \right]}-\frac{\partial W}{\partial {{\left( {{{\mathbf{\dot{u}}}}^{e}} \right)}_{v}}}=0,
\label{13}
\end{aligned}
\end{equation}

\begin{equation}
\begin{aligned}
\frac{\delta {{E}_{K}}}{\delta {{\left( {{\mathbf{M}}^{q}} \right)}_{v}}}-\frac{\partial \Phi }{\partial {{\left( {{\mathbf{M}}^{q}} \right)}_{v}}}+\sum\limits_{i}{\frac{d}{d{{r}_{i}}}\left[ \frac{\partial \Phi }{\partial {{\left( \mathbf{M}_{,i}^{q} \right)}_{v}}} \right]}-\frac{\partial W}{\partial {{\left( {{{\mathbf{\dot{M}}}}^{q}} \right)}_{v}}}=0.
\label{14}
\end{aligned}
\end{equation}
Generally speaking, eqs. (11-14) are difficult to solve, since the presence of ${{\left( {{\mathbf{u}}^{e}}(\mathbf{r},t) \right)}_{v}}$ and ${{\left( {{\mathbf{M}}^{q}}(\mathbf{r},t) \right)}_{v}}$ in eq. (8) breaks the orthogonality of the Fourier series expression of $\mathbf{M}$ introduced in eq. (7). Hereafter we focus on the solution of eqs. (11-12) for small amplitude vibration of ${{\left( {{\mathbf{u}}^{e}}(\mathbf{r},t) \right)}_{v}}$ and ${{\left( {{\mathbf{M}}^{q}}(\mathbf{r},t) \right)}_{v}}$ at long wavelength limit, while eqs. (13-14) can be treated in a similar way.
Consider the emergent elastic wave propagation in the EC with small amplitude, i.e., ${{\left( {{\mathbf{u}}^{e}}(\mathbf{r},t) \right)}_{v}}$ and ${{\left( {{\mathbf{M}}^{q}}(\mathbf{r},t) \right)}_{v}}$ have small magnitudes and they both change smoothly in space. In this case, the orthogonality of the Fourier series expression of $\mathbf{M}$ is approximately maintained, and we can expand $\Phi $ as
\begin{widetext}
\begin{equation}
\begin{aligned}
  & \Phi ={{\left( \Phi  \right)}_{st}}+\frac{1}{2}{{\sum\limits_{i,j}{\left( \frac{{{\partial }^{2}}\Phi }{\partial {{\left( u_{i}^{e} \right)}_{v}}\partial {{\left( u_{j}^{e} \right)}_{v}}} \right)}}_{st}}{{\left( u_{i}^{e} \right)}_{v}}{{\left( u_{j}^{e} \right)}_{v}}+\frac{1}{2}{{\sum\limits_{i,j,k,l}{\left( \frac{{{\partial }^{2}}\Phi }{\partial {{\left( u_{i,k}^{e} \right)}_{v}}\partial {{\left( u_{j,l}^{e} \right)}_{v}}} \right)}}_{st}}{{\left( u_{i,k}^{e} \right)}_{v}}{{\left( u_{j,l}^{e} \right)}_{v}} \\ 
 & +{{\sum\limits_{i,j,k}{\left( \frac{{{\partial }^{2}}\Phi }{\partial {{\left( u_{i,k}^{e} \right)}_{v}}\partial {{\left( u_{j}^{e} \right)}_{v}}} \right)}}_{st}}{{\left( u_{i,k}^{e} \right)}_{v}}{{\left( u_{j}^{e} \right)}_{v}}+\frac{1}{2}{{\sum\limits_{i,j}{\left( \frac{{{\partial }^{2}}\Phi }{\partial {{\left( M_{i}^{q} \right)}_{v}}\partial {{\left( M_{j}^{q} \right)}_{v}}} \right)}}_{st}}{{\left( M_{i}^{q} \right)}_{v}}{{\left( M_{j}^{q} \right)}_{v}} \\ 
 & +\frac{1}{2}{{\sum\limits_{i,j,k,l}{\left( \frac{{{\partial }^{2}}\Phi }{\partial {{\left( M_{i,k}^{q} \right)}_{v}}\partial {{\left( M_{j,l}^{q} \right)}_{v}}} \right)}}_{st}}{{\left( M_{i,k}^{q} \right)}_{v}}{{\left( M_{j,l}^{q} \right)}_{v}}+{{\sum\limits_{i,j,k}{\left( \frac{{{\partial }^{2}}\Phi }{\partial {{\left( M_{i,k}^{q} \right)}_{v}}\partial {{\left( M_{j}^{q} \right)}_{v}}} \right)}}_{st}}{{\left( M_{i,k}^{q} \right)}_{v}}{{\left( M_{j}^{q} \right)}_{v}} \\ 
 & +{{\sum\limits_{i,j,k}{\left( \frac{{{\partial }^{2}}\Phi }{\partial {{\left( u_{i,k}^{e} \right)}_{v}}\partial {{\left( M_{j}^{q} \right)}_{v}}} \right)}}_{st}}{{\left( u_{i,k}^{e} \right)}_{v}}{{\left( M_{j}^{q} \right)}_{v}}+{{\sum\limits_{i,j,k}{\left( \frac{{{\partial }^{2}}\Phi }{\partial {{\left( M_{i,k}^{q} \right)}_{v}}\partial {{\left( u_{j}^{e} \right)}_{v}}} \right)}}_{st}}{{\left( M_{i,k}^{q} \right)}_{v}}{{\left( u_{j}^{e} \right)}_{v}} \\ 
 & +{{\sum\limits_{i,j}{\left( \frac{{{\partial }^{2}}\Phi }{\partial {{\left( u_{i}^{e} \right)}_{v}}\partial {{\left( M_{j}^{q} \right)}_{v}}} \right)}}_{st}}{{\left( u_{i}^{e} \right)}_{v}}{{\left( M_{j}^{q} \right)}_{v}}+{{\sum\limits_{i,j,k,l}{\left( \frac{{{\partial }^{2}}\Phi }{\partial {{\left( u_{i,k}^{e} \right)}_{v}}\partial {{\left( M_{j,l}^{q} \right)}_{v}}} \right)}}_{st}}{{\left( u_{i,k}^{e} \right)}_{v}}{{\left( M_{j,l}^{q} \right)}_{v}}. 
\label{15}
\end{aligned}
\end{equation}
The last term on the r.h.s. of eq. (15) is included when an AC magnetic field is applied to the material. For the kinetic energy of the system, we have
\begin{equation}
\begin{aligned}
\frac{\delta {{E}_{K}}}{\delta {{\left( u_{i}^{e} \right)}_{v}}}=\sum\limits_{j}{{{\left( \frac{\partial }{\partial {{\left( \dot{u}_{j}^{e} \right)}_{v}}}\frac{\delta {{E}_{K}}}{\delta {{\left( u_{i}^{e} \right)}_{v}}} \right)}_{st}}{{\left( \dot{u}_{j}^{e} \right)}_{v}}}+\sum\limits_{j}{{{\left( \frac{\partial }{\partial {{\left( \dot{M}_{j}^{q} \right)}_{v}}}\frac{\delta {{E}_{K}}}{\delta {{\left( u_{i}^{e} \right)}_{v}}} \right)}_{st}}{{\left( \dot{M}_{j}^{q} \right)}_{v}}},
\label{16}
\end{aligned}
\end{equation}
\begin{equation}
\begin{aligned}
\frac{\delta {{E}_{K}}}{\delta {{\left( M_{i}^{q} \right)}_{v}}}=\sum\limits_{j}{{{\left( \frac{\partial }{\partial {{\left( \dot{u}_{j}^{e} \right)}_{v}}}\frac{\delta {{E}_{K}}}{\delta {{\left( M_{i}^{q} \right)}_{v}}} \right)}_{st}}{{\left( \dot{u}_{j}^{e} \right)}_{v}}}+\sum\limits_{j}{{{\left( \frac{\partial }{\partial {{\left( \dot{M}_{j}^{q} \right)}_{v}}}\frac{\delta {{E}_{K}}}{\delta {{\left( M_{i}^{q} \right)}_{v}}} \right)}_{st}}{{\left( \dot{M}_{j}^{q} \right)}_{v}}},
\label{17}
\end{aligned}
\end{equation}
\end{widetext}
Substitution of eqs. (15-17) into eqs. (11, 12) yields a set of linearized partial differential equations for ${{\left( {{\mathbf{u}}^{e}}(\mathbf{r},t) \right)}_{v}}$ and ${{\left( {{\mathbf{M}}^{q}}(\mathbf{r},t) \right)}_{v}}$. Consider the plane-wave form of solution ${{\left( {{\mathbf{u}}^{e}}(\mathbf{r},t) \right)}_{v}}={{\mathbf{u}}^{e0}}{{e}^{\text{i}(\mathbf{\tilde{k}}\cdot \mathbf{r}-\omega t)}}$, ${{\left( {{\mathbf{M}}^{q}}(\mathbf{r},t) \right)}_{v}}={{M}^{q0}}{{e}^{\text{i}(\mathbf{\tilde{k}}\cdot \mathbf{r}-\omega t)}}$, a generalized eigenvalue problem of the frequency $\omega $ can be obtained as
\begin{equation}
\begin{aligned}
\left( \mathbf{R}\omega -\mathbf{K} \right)\left[ \begin{matrix}
   {{\mathbf{u}}^{e0}}  \\
   {{\mathbf{M}}^{q0}}  \\
\end{matrix} \right]=\mathbf{0},
\label{18}
\end{aligned}
\end{equation}
where
\begin{equation}
\begin{aligned}
\mathbf{R}=\left[ \begin{matrix}
   {{\mathbf{R}}^{e}} & {{\mathbf{R}}^{eq}}  \\
   {{\left( {{\mathbf{R}}^{eq*}} \right)}^{T}} & {{\mathbf{R}}^{q}}  \\
\end{matrix} \right],
\label{19}
\end{aligned}
\end{equation}

\begin{equation}
\begin{aligned}
\mathbf{K }=\left[ \begin{matrix}
   {{\mathbf{K }}^{e}} & {{\mathbf{K }}^{eq}}  \\
   {{\left( {{\mathbf{K }}^{eq*}} \right)}^{T}} & {{\mathbf{K }}^{q}}  \\
\end{matrix} \right],
\label{20}
\end{aligned}
\end{equation}
where ${{\mathbf{R}}^{eq*}}$ and ${{\mathbf{K }}^{eq*}}$ denote complex conjugate of ${{\mathbf{R}}^{eq}}$ and ${{\mathbf{K}}^{eq}}$. $R_{ij}^{e}=-\text{i}{{\left[ \frac{\partial }{\partial \dot{u}_{j}^{e}}\left( \frac{\delta {{E}_{BP}}}{\delta u_{i}^{e}} \right) \right]}_{st}},$ $R_{ij}^{q}= -\text{i}{{\left[ \frac{\partial }{\partial \dot{M}_{j}^{q}}\left( \frac{\delta {{E}_{BP}}}{\delta M_{i}^{q}} \right) \right]}_{st}},$ $R_{ij}^{eq}= -\text{i}{{\left[ \frac{\partial }{\partial \dot{M}_{j}^{q}}\left( \frac{\delta {{E}_{BP}}}{\delta u_{i}^{e}} \right) \right]}_{st}}$, $K _{ij}^{e}=\sum\limits_{p,s}{{{{\tilde{k}}}_{p}}{{{\tilde{k}}}_{s}}{{\left[ \frac{\partial }{\partial u_{j,ps}^{e}}\left( \frac{d}{d{{r}_{p}}}\left( \frac{\partial \bar{\Phi }}{\partial u_{i,p}^{e}} \right) \right) \right]}_{st}}},$ $K _{ij}^{eq}={{\left[ \sum\limits_{p,s}{{{{\tilde{k}}}_{p}}{{{\tilde{k}}}_{s}}}\frac{\partial }{\partial M_{j,ps}^{q}}\left( \frac{d}{d{{r}_{p}}}\frac{\partial \Phi }{\partial u_{i,p}^{e}} \right)-\sum\limits_{p}{\text{i}}{{{\tilde{k}}}_{p}}\frac{\partial }{\partial M_{j,p}^{q}}\left( \frac{d}{d{{r}_{p}}}\frac{\partial \Phi }{\partial u_{i,p}^{e}} \right) \right]}_{st}},$ $ K _{ij}^{q}={{\left[ \frac{\partial }{\partial M_{j}^{q}}\left( \frac{\partial \bar{\Phi }}{\partial M_{i}^{q}} \right)+\sum\limits_{p}{\text{i}}{{{\tilde{k}}}_{p}}\frac{\partial }{\partial M_{j,p}^{q}}\left( \frac{\partial \Phi }{\partial M_{i}^{q}} \right)-\sum\limits_{p}{\text{i}}{{{\tilde{k}}}_{p}}\frac{\partial }{\partial M_{j,p}^{q}}\right.}}$ ${{\left.\left( \frac{d}{d{{r}_{p}}}\left( \frac{\partial \Phi }{\partial M_{i,p}^{q}} \right) \right)+\sum\limits_{p,s}{{{{\tilde{k}}}_{p}}{{{\tilde{k}}}_{s}}}\frac{\partial }{\partial M_{j,ps}^{q}}\left( \frac{d}{d{{r}_{p}}}\left( \frac{\partial \Phi }{\partial M_{i,p}^{q}} \right) \right) \right]}_{st}}.$ Here a subscript $"st"$ means that the term is calculated at the equilibrium state ${{\mathbf{u}}^{e}}={{\left( {{\mathbf{u}}^{e}} \right)}_{st}}$ and ${{\mathbf{M}}^{q}}={{\left( {{\mathbf{M}}^{q}} \right)}_{st}}$.

When $\mathbf{\tilde{k}}\to 0$, the “stiffness matrix” $\mathbf{K}$ is completely determined by the emergent elastic properties of the EC considered\cite{11}. To be more specific ${{\mathbf{K}}^{q}}={{\mathbf{\mu }}^{q}}$, ${{\mathbf{K}}^{e}}$ is determined by ${{\mathbf{C}}^{e}}$ thorough
\begin{widetext}
\begin{equation}
\begin{aligned}
& K_{11}^{e}=-C_{11}^{e}\tilde{k}_{1}^{2}-\frac{1}{4}\left( C_{33}^{e}+C_{44}^{e}+2C_{34}^{e} \right)\tilde{k}_{2}^{2}-\left( C_{13}^{e}+C_{14}^{e} \right){{{\tilde{k}}}_{1}}{{{\tilde{k}}}_{2}}, \\ 
 & K_{22}^{e}=-C_{22}^{e}\tilde{k}_{2}^{2}-\frac{1}{4}\left( C_{33}^{e}+C_{44}^{e}-2C_{34}^{e} \right)\tilde{k}_{1}^{2}-\left( C_{23}^{e}-C_{24}^{e} \right){{{\tilde{k}}}_{1}}{{{\tilde{k}}}_{2}}, \\ 
 & K_{12}^{e}=K_{21}^{e}=-\frac{1}{2}\left( C_{13}^{e}-C_{14}^{e} \right)\tilde{k}_{1}^{2}-\frac{1}{2}\left( C_{23}^{e}+C_{24}^{e} \right)\tilde{k}_{2}^{2}-\frac{1}{2}\left( 2C_{12}^{e}+C_{33}^{e}-C_{44}^{e} \right){{{\tilde{k}}}_{1}}{{{\tilde{k}}}_{2}}, \\
 \label{21} 
\end{aligned}
\end{equation}
\end{widetext}
and ${{\mathbf{K}}^{eq}}$ is determined by ${{\mathbf{g}}^{eq}}$ through 
\begin{equation}
\begin{aligned}
  & K_{1i}^{eq}=\text{i}\left( g_{1i}^{eq}{{{\tilde{k}}}_{1}}+\frac{1}{2}g_{3i}^{eq}{{{\tilde{k}}}_{2}}+\frac{1}{2}g_{4i}^{eq}{{{\tilde{k}}}_{2}} \right), \\ 
 & K_{2i}^{eq}=\text{i}\left( g_{2i}^{eq}{{{\tilde{k}}}_{1}}+\frac{1}{2}g_{3i}^{eq}{{{\tilde{k}}}_{2}}-\frac{1}{2}g_{4i}^{eq}{{{\tilde{k}}}_{2}} \right). \\
\label{22}
\end{aligned}
\end{equation}
Based on the formulation established here, we systematically study the long wavelength emergent phonon excitations for isotropic hexagonal SkX\cite{29} and distorted SkX due to presence of anisotropic effects\cite{30}.

\begin{acknowledgments}
The author gratefully acknowledges J. D. Zang for helpful discussion. The work was supported by the NSFC (National Natural Science Foundation of China) through the funds 11772360 and Pearl River Nova Program of Guangzhou (Grant No. 201806010134).
\end{acknowledgments}

\bibliographystyle{apsrev4-1}

\begin{thebibliography}{99}
   
    \bibitem{1} 	Mühlbauer, S. et al. Skyrmion Lattice in a Chiral Magnet. Science \textbf{323} 915–919 (2009).
    \bibitem{2} 	Kezsmarki, I. et al. Neel-type skyrmion lattice with confined orientation in the polar magnetic semiconductor GaV4S8. Nature Materials \textbf{14} 1116-+ (2015).
    \bibitem{3} 	von Bergmann, K., Menzel, M., Kubetzka, A. \& Wiesendanger, R. Influence of the Local Atom Configuration on a Hexagonal Skyrmion Lattice. Nano Letters \textbf{15} 3280–3285 (2015).
    \bibitem{4} 	Langner, M. C. et al. Coupled Skyrmion Sublattices in Cu2OSeO3. Physical Review Letters \textbf{112} 167202 (2014).
    \bibitem{5} 	Tanigaki, T. et al. Real-Space Observation of Short-Period Cubic Lattice of Skyrmions in MnGe. Nano Letters \textbf{15} 5438–5442 (2015).
    \bibitem{6} 	Nayak, A. K. et al. Magnetic antiskyrmions above room temperature in tetragonal Heusler materials. Nature \textbf{548} 561–566 (2017).
    \bibitem{7} 	Yu, X. Z. et al. Biskyrmion states and their current-driven motion in a layered manganite. Nature Communications \textbf{5} 3198 (2014).
    \bibitem{8} 	Wang, W. et al. A Centrosymmetric Hexagonal Magnet with Superstable Biskyrmion Magnetic Nanodomains in a Wide Temperature Range of 100–340 K. Advanced Materials \textbf{28} 6887–6893 (2016).
    \bibitem{9} 	Onose, Y., Okamura, Y., Seki, S., Ishiwata, S. \& Tokura, Y. Observation of Magnetic Excitations of Skyrmion Crystal in a Helimagnetic Insulator Cu2OSeO3. Physical Review Letters \textbf{109} 037603 (2012).
    \bibitem{10} 	Schwarze, T. et al. Universal helimagnon and skyrmion excitations in metallic, semiconducting and insulating chiral magnets. Nature Materials \textbf{14} 478–483 (2015).
    \bibitem{11} 	Buettner, F. et al. Dynamics and inertia of skyrmionic spin structures. Nature Physics \textbf{11} 225–228 (2015).
    \bibitem{12} 	Cote, R., Luo, W., Petrov, B., Barlas, Y. \& MacDonald, A. H. Orbital and interlayer skyrmion crystals in bilayer graphene. Physical Review B \textbf{82} 245307 (2010).
    \bibitem{13} 	Garst, M., Waizner, J. \& Grundler, D. Collective spin excitations of helices and magnetic skyrmions: review and perspectives of magnonics in non-centrosymmetric magnets. Journal of Physics D-Applied Physics \textbf{50} 293002 (2017).
    \bibitem{14} 	Roldán-Molina, A., Nunez, A. S. \& Fernández-Rossier, J. Topological spin waves in the atomic-scale magnetic skyrmion crystal. New J. Phys. \textbf{18} 045015 (2016).
    \bibitem{15} 	Takagi, R. et al. Spin-wave spectroscopy of the Dzyaloshinskii-Moriya interaction in room-temperature chiral magnets hosting skyrmions. Physical Review B \textbf{95} 220406 (2017).
    \bibitem{16} 	Mochizuki, M. Spin-Wave Modes and Their Intense Excitation Effects in Skyrmion Crystals. Physical Review Letters \textbf{108} 017601 (2012).
    \bibitem{17} 	Petrova, O. \& Tchernyshyov, O. Spin waves in a skyrmion crystal. Phys. Rev. B \textbf{84} 214433 (2011).
    \bibitem{18} 	Tucker, G. S. et al. Spin excitations in the skyrmion host Cu2OSeO3. Physical Review B \textbf{93} 054401 (2016).
    \bibitem{19} 	Langner, M. C. et al. Nonlinear Ultrafast Spin Scattering in the Skyrmion Phase of Cu2OSeO3. Physical Review Letters \textbf{119} 107204 (2017).
    \bibitem{20} 	Mochizuki, M. \& Seki, S. Magnetoelectric resonances and predicted microwave diode effect of the skyrmion crystal in a multiferroic chiral-lattice magnet. Physical Review B \textbf{87} (2013).
    \bibitem{21} 	Mruczkiewicz, M., Gruszecki, P., Zelent, M. \& Krawczyk, M. Collective dynamical skyrmion excitations in a magnonic crystal. Physical Review B \textbf{93} 174429 (2016).
    \bibitem{22} 	Landau, L. D. \& Lifshitz, E. M. Theory of the dispersion of magnetic permeability in ferromagnetic bodies. Phys. Z. Sowietunion \textbf{8} 153–169 (1935).
    \bibitem{23} 	Gilbert, T. L. Lagrangian formulation of the gyromagnetic equation of the magnetization field. Phys. Rev. \textbf{100} 1243–1243 (1955).
    \bibitem{24} 	Hu, Y. Wave nature and metastability of emergent crystals in chiral magnets. Communications Physics \textbf{1} 82 (2018).
    \bibitem{25} 	Hu, Y. \& Wan, X. Thermodynamics and elasticity of emergent crystals. arXiv: 1905.02165 (2019).
    \bibitem{26} 	Zang, J., Mostovoy, M., Han, J. H. \& Nagaosa, N. Dynamics of Skyrmion Crystals in Metallic Thin Films. Phys. Rev. Lett. \textbf{107} 136804 (2011).
    \bibitem{27} 	Tatara, G., Kohno, H. \& Shibata, J. Microscopic approach to current-driven domain wall dynamics. Physics Reports \textbf{468} 213–301 (2008).
    \bibitem{28} 	Brown, W. F. Jr. Micromagnetics. (Interscience Tracts on Physics and Astronomy Publishers, 1963).
    \bibitem{29} 	Hu, Y. Emergent elastic waves in skyrmion crystals with finite frequencies at long wavelength limit. ArXiv (2019).
    \bibitem{30} 	Hu, Y. Long-wavelength emergent phonons in distorted skyrmion crystals induced by intrinsic exchange anisotropy and a tilted magnetic field. ArXiv (2019).
    

\end{thebibliography}

\end{document}